\documentstyle[11pt,paspconf,epsf]{article}

\def\etal{{\sl et al.}}

\def\kms{km s$^{-1}$}

\def\ergs{ergs s$^{-1}$}

\def\lsun{L$_{\odot}$}

\def\halpha{\ifmmode {\rm H{\alpha}} \else $\rm H{\alpha}$\fi}
\def\hbeta{\ifmmode {\rm H{\beta}} \else $\rm H{\beta}$\fi}
\def\heiia{He\,{\sc ii} $\lambda$4686}

\def\niii{N\,{\sc iii} $\lambda$4640}

\def\nvb{N\,{\sc v} $\lambda$1240}
\def\ciii{C\,{\sc iii} $\lambda$5696}
\def\civ{C\,{\sc iv} $\lambda$5808}
\def\civb{C\,{\sc iv} $\lambda$1549}
\def\siiv{[Si\,{\sc iv}] $\lambda$1400}
\begin{document}

{\footnotesize \it
\noindent
Wolf-Rayet Phenomena in Massive Stars and Starburst Galaxies \\
Proceedings IAU Symposium No. 193, \copyright 1999 IAU \\
K.A. van der Hucht, G. Koenigsberger \& P.R.J. Eeenens, eds.
}

\medskip
\title{Wolf-Rayet stars as tracing the AGN-starburst connection}
\author{Daniel Kunth}
\affil{Institut d'Astrophysique de Paris, France}
\author{Thierry Contini}
\affil{European Southern Observatory, Garching, Germany}



\begin{abstract}
{\small
We stress the importance of Wolf-Rayet stars for the understanding 
of the AGN phenomenon in galaxies. WR stars provide an unique opportunity 
to explore from the ground whether non-thermal nuclear activity and 
circumnuclear starbursts are connected. We review the known reported 
WR signatures observed so far in AGNs and point out some intrincacies 
related to the analysis of the spectra, linked to reddening correction, 
the origin of the \hbeta\ line, etc. Finally, we advocate that integral 
field spectroscopy is a very promising tool to study this problem and present 
preliminary results of a long-term project that have been obtained at the 
{\sl CFHT} in 1998.
}
\end{abstract}

\section{Introduction}

The nature of the dominant ionizing source in Active Galactic Nuclei (AGNs) 
is still poorly understood.
Under the unified scenario, Seyfert 1 and 
Seyfert 2 galaxies represent the same kind of objects seen from different 
angles by the observer. In Seyfert 1 galaxies, one sees the central engine 
(a black hole surrounded by an accretion disc) whereas in Seyfert 2 galaxies 
a dusty torus completely blocks the non-thermal phenomenon. 
On the other hand, it has been suggested that starbursts may play an important 
role in the Seyfert phenomenon. In this case, AGNs are 
various distinct evolutionary phases of a burst of star formation in the core 
of early-type spiral galaxies. Physical processes associated with the starburst 
phenomenon can in principle account for most of the AGNs properties. In an
hybrid 
version that take advantage of both views, a very compact circumnuclear
starburst 
coexists with a massive central black hole responsible for the ionizing
continuum. 
But a fundamental question remains to be answered. 
Are the non-thermal nuclear activity and the circumnuclear starbursts connected?

The question of the origin of the ultraviolet (UV) continuum in Seyfert 2
galaxies
has recently received new attention. 
The advantage in chosing Seyfert 2 galaxies for further investigation of the 
AGN-starburst phenomenon is that one is not hampered by the central source.
Early studies of {\sl IUE} spectra by Heckman \etal\ (1995) showed that the UV
radiation of
Seyfert 2 galaxies is not compatible with the scenario of radiation scattered
from
an obscured nucleus. Instead its shape was fully understandable as being
produced by a reddened starburst. {\sl HST} has
revealed that the UV continuum is spatially-resolved. Moreover, four bright
Seyfert 2 galaxies have spectra dominated by the clear spectroscopic signatures
of a starburst population. Strong stellar wind features were indeed detected
(\nvb, \siiv, and \civb) as well as stellar photospheric lines (Heckman \etal\
1997, 
Gonzalez Delgado \etal\ 1999). In these four Seyfert 2 galaxies, 
selected on the basis of their UV brigthness, the data provide evidence of the 
existence of nuclear starbursts. These account for a large fraction of the
total 
intrinsic luminosity of the entire galaxy.

\section{The Wolf-Rayet population in emission-line galaxies}

\subsection{WR signatures in AGNs}
UV studies have open new ways of studying the possible presence of a powerful 
starburst in AGNs. However, UV--bright Seyfert 2 galaxies are not so numerous, 
mostly because of dust obscuration, hence detailed UV observations are
difficult 
to achieve even with the {\sl HST}. This is why we believe that Wolf-Rayet (WR) stars 
are excellent 
tracers to follow up in relation with the AGN-starburst phenomenon, 
mainly because they provide, in turn, conspicuous signatures in the visible. 
The detection of the broad emission feature at \heiia\ is indeed 
attributed mainly to WR stars of WN subtype, whereas \civ, the strongest 
emission line of WC stars, is now currently observed in WR galaxies 
(Schaerer, Contini \& Kunth 1999). 
Moreover, although they represent only a special short-lived phase in the 
evolution of the starburst population,  
their occurence is highly expected in metal--rich environment. 

Mrk 477 is the best case of Seyfert 2 galaxy where the possible signature 
of massive WR stars has been detected (Heckman \etal\ 1997). 
But, because the broad stellar feature around 4686 \AA\ is very faint, 
it is difficult to separate this component from the bright \heiia\ 
nebular emission-line. 
The narrow nebular contribution can in principle be removed but 
illustrate the difficulty of detecting the WR features around 4700 \AA. 
For this reason, the WR feature at \civ\ could be a better tool to detect 
massive stars in AGNs because it is free from unwanted nebular contributions 
(see sect.~\ref{results}). 
Mrk 477 has a very blue UV continuum, and a compact dusty starburst of a few 
hundreds parsecs dominates the light from the UV to the near infrared. 
The bolometric luminosity of this nuclear starburst is $4 \times 10{^4}$ \lsun, 
a significant fraction of the total bolometric luminosity of Mrk 477.
In the other three Seyferts (NGC 7130, NGC 5135 and IC 3639) studied by
Gonzalez 
Delgado \etal\ (1999), \heiia\ seems to be of nebular origin.

\subsection{WR Galaxies, IRAS starburts and AGNs}

As pointed out by Contini (these Proceedings), about 140 WR galaxies are now
cataloged in the last compilation of Schaerer, Contini \& Pindao (1999). 
From previous detailed studies of WR galaxies some complications have been
identified that one must keep in mind for AGN studies. 

One key issue is the
method that one adopts to estimate the number of WR stars. A convenient way  to 
quantify the WR and the O stellar populations is to use the \heiia/\hbeta\
luminosity 
ratios (Kunth \& Sargent 1981; Vacca \& Conti 1992; Vacca 1994). 
This method has some 
possible drawbacks (see also Schaerer, these Proceedings). In Table 1, we report 
some results that have been derived for four types of galaxies: 
Mrk 477 (Seyfert 2), Mrk 309 
(a LINER previously cataloged as Seyfert 2, then as a WR galaxy), NGC6764 
(first known as a Seyfert 2, then a WR galaxy and now reported as a LINER) and 
IRAS 01003-2238, a massive dusty starburst galaxy. 
One may ask whether the WR phenomenon in AGNs is a scaled up version of what 
we observe in intrinsically fainter WR galaxies. Indeed as shown by Kunth \&
Schild 
(1986, hereafter KS86), one finds that as the absolute luminosity of galaxies 
increases so does the metallicity of the galaxy hence the luminosity $L_{\rm
WR}$ 
of the WR feature. 
Using the Table 1, we can check that $L_{\rm WR}$ nicely falls onto the right 
side of KS86's fig. 5 where they plot metal-rich galaxies with the strongest 
$L_{\rm WR}$ luminosities. 
How well estimated
are the WR/O ratios? KS86 argue that if one uses E($B-V$)
derived from the Balmer decrement one makes the implicit assumption that all 
massive stars are deeply nested inside the nebular complex. Mas-Hesse \& Kunth 
(1999) have shown that the extinction obtained from the Balmer decrement is
larger 
than the one derived from the UV continuum. This imply that the WR
feature should be dereddened with a different correction value than the 
\hbeta\ line and lead to a much smaller WR number than what is reported in
Table 1. 
Another concern is the use of the \hbeta\ line which is also produced by the 
non-thermal central source and not only by high-mass stars as pointed out by 
Heckman \etal\ (1997) ; explaining why the WR/\hbeta\ flux ratio is not as 
large in Mrk 477 than in the less active galaxies. 

\begin{table}
\caption{WR population in massive emission-line galaxies}
{\footnotesize
\begin{tabular}{lccccc}
\hline
\hline
galaxy    & Mrk~477 &  Mrk~309$^1$ & NGC~6764$^1$ & IRAS~01003-2238$^2$   \\
\hline 
morphology & compact & Sa & SBbc & ... \\
$v_{\rm rad}$ (\kms)  & 11494 & 12831 & 2637 & 35328 \\
$M_{\rm abs}$ & --21.0 & --21.7 & --21.1 & ... \\ 
\heiia/\hbeta & 0.10 & 0.67 & 0.43 & 0.61 \\
EW(\heiia) & 12\AA\ & 8\AA\ & 9\AA\ & ... \\
E($B-V$) & 0.23 & 0.7 & ... & ... \\
$L_{\rm WR}$ (\ergs) & $5 \times 10^{40}$ & $5 \times 10^{41}$ & ... & $3
\times 10^{41}$ \\
$N_{\rm WR}$ & $3 \times 10^{4}$ & $>10^{5}$ & $3 \times 10^{3}$ & $10^{5}$ \\
WR/O & 0.4 & ... & ... & ... \\
$12 + \log(\rm O/H)$  & ... & 8.90 & 8.88 & ... \\
\hline
\hline
\multicolumn{5}{c}{References to Table 1: (1) Osterbrock \& Cohen (1982); (2)
Armus et al. (1988)} \\
\end{tabular}
}
\label{DATA}
\end{table}

\section{Integral field spectroscopy of nearby Seyfert 2 galaxies}

\subsection{Motivation}

We started an long-term project to clearly establish a direct link 
between the population of massive stars formed during a starburst and 
the AGN phenomenon. Indeed, despite the promising results obtained by Heckman 
and his collaborators (these Proceedings), fundamental questions remain. 
Is Mrk 477 an exception among Seyfert 2 galaxies? Where is precisely 
located the starburst detected by Heckman \etal\ (1997) in Mrk 477: 
exactly in the nucleus or in the circumnuclear region? Indeed, because 
of its high distance ($\sim$ 153 Mpc), Heckman \etal\ were not able to 
locate the massive stars at a resolution better than $\sim$ 1 kpc, 
which is insufficient to assert that the starburst is nuclear.

The main goal of our program is to detect the spectral signatures 
of an important population of massive WR stars in the nucleus of 
Seyfert 2 galaxies. From the lists of Heckman \etal\ (1995), 
Terlevich \etal\ (1990), and Gonzalez Delgado \etal\ (1997), 
we selected a dozen of Seyfert 2 galaxies with very blue spectra. 
The advantage of this sample is to focus only on the 
nearest AGNs, located at a maximum distance of $\sim$ 60 Mpc 
(with H$_0 = 75$ kms$^{-1}$Mpc$^{-1}$). 
Using MOS-ARGUS on the 3.6m {\sl CFHT} with a FoV of $\sim 12\arcsec 
\times 7\arcsec$ and 
a spatial sampling of 0.4\arcsec/fiber, it is thus possible to explore 
the central region of galaxies with a resolution of $\sim$ 100 pc 
(assuming a seeing of 0.6\arcsec).  

\subsection{Preliminary results}
\label{results}

Bi-dimensional spectroscopic observations of 5 galaxies (see Table~\ref{SEY2}) 
were obtained on the nights of 1998 
May 31 -- June 2 at {\sl CFHT}. 
The data were acquired with ARGUS, a special mode for integral field 
spectroscopy with MOS. We used the blue grism B600 which gives 
a spectral coverage of about 3600 -- 7200 \AA\ with a resolution 
of $\sim$ 4 \AA. The total integration time for each galaxy ranges 
from 90 to 150 min in order to achieve a S/N greater than 30 around 
4700 and 5800 \AA. The data reduction is in progress but preliminary 
results already illustrate the feasibility of this program. 

\begin{figure}
\plottwo{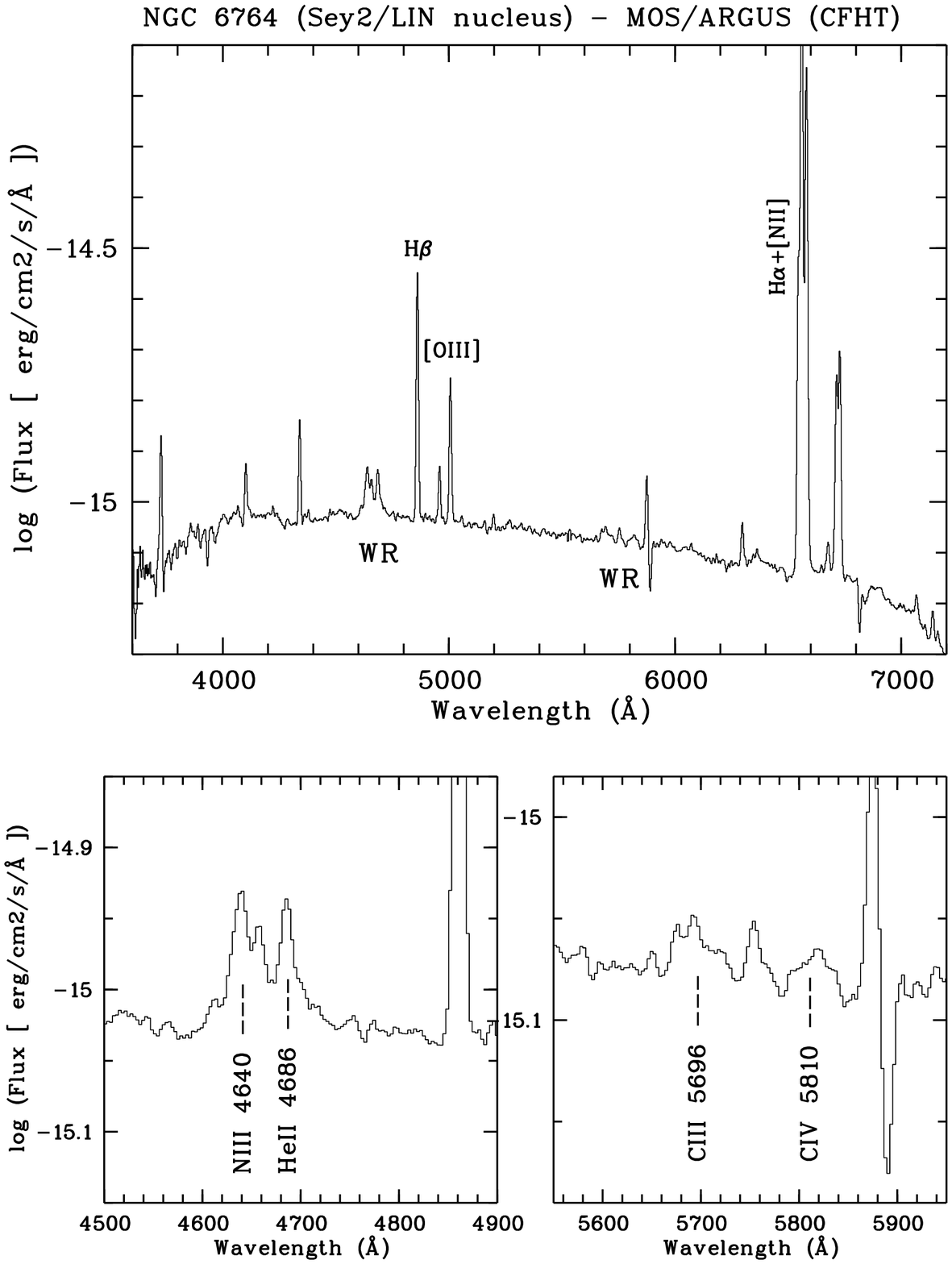}{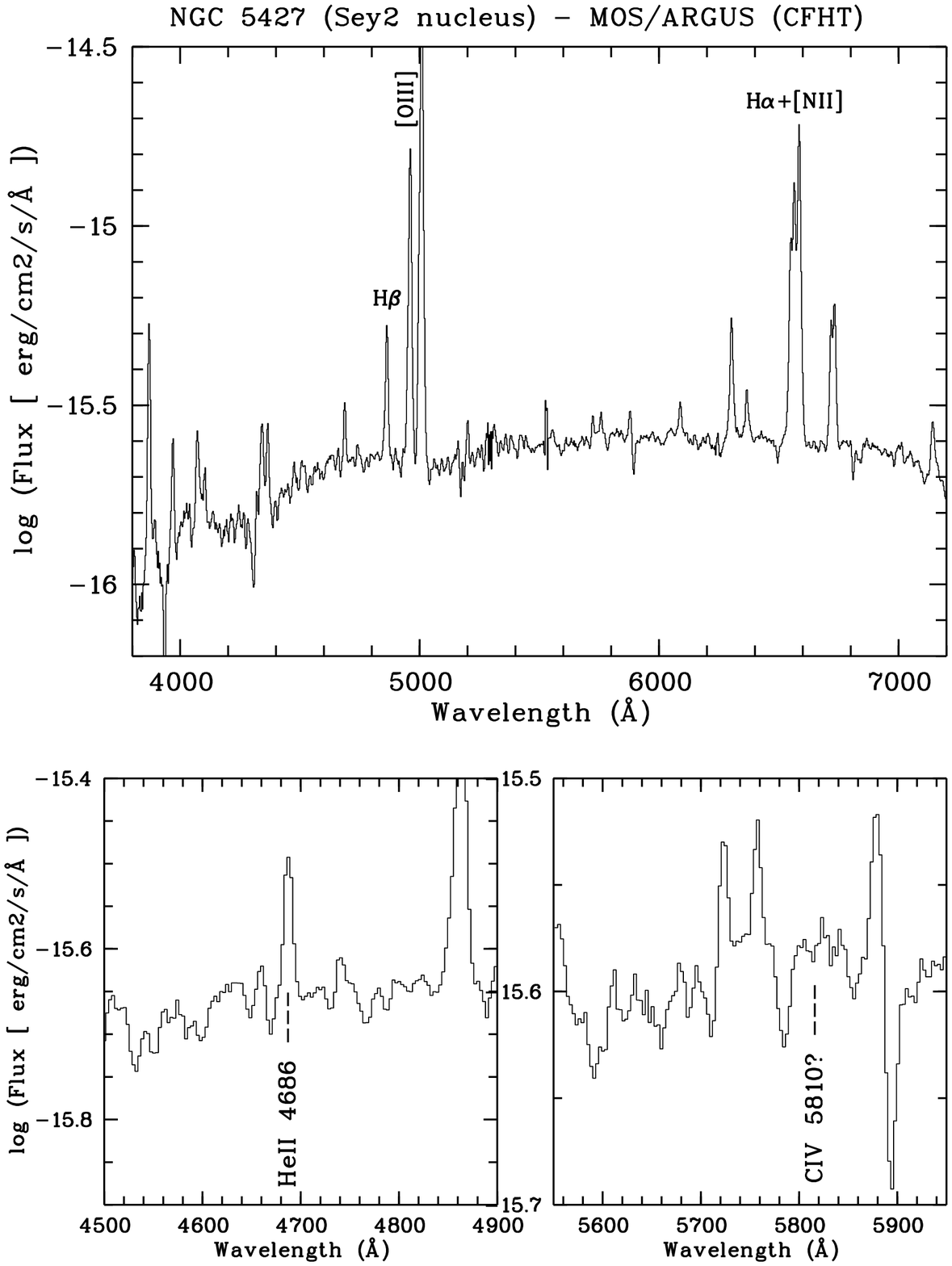}
\vspace{-1cm}
\caption{\small Nuclear spectra of two nearby Seyfert 2 galaxies observed with 
MOS/ARGUS at {\sl CFHT}. NGC 6764 (left) represents the best case of an AGN 
where the spectral signatures of both WN (\niii\ and \heiia) and 
WC stars (\ciii\ and \civ) are clearly detected, indicating a young 
and massive starburst in the core of this AGN. The detection of 
WR stars in the nucleus of NGC 5427 (right) is less clear. The 
blue WR bump is contaminated by the bright nebular \heiia\ 
emission line and the S/N in the red part of the spectrum is to 
low to see clearly the \civ\ line.}
\label{SPEC}
\end{figure}

In Fig.~\ref{SPEC}, we show the 
nuclear spectra of NGC 6764 and NGC 5427, two Seyfert 2 galaxies 
where we detected the spectral signature of WR stars. 
NGC 6764 
is the best example in which broad emission lines from both 
WN and WC stars are clearly observed. Whereas the blue WR bump 
(\niii\ and \heiia) was already known from previous long-slit 
observations (Osterbrock \& Cohen 1982, Eckart \etal\ 1996), 
we detect for the first time WR features in the red part of 
the spectrum (\ciii\ and \civ), indicating an important 
population of massive WC stars in this object. Moreover, 
the \ciii\ broad emission line, which is rarely observed in WR 
galaxies (only two cases so far: NGC 1365, Phillips \& Conti 1992; 
and NGC 3049, Schaerer, Contini \& Kunth 1999), indicates 
a substantial population of late-type WC stars as expected 
in high-metallicity environment.
As can be seen in Fig.~\ref{SPEC}, the blue WR bump around 
4700 \AA\ is usually contaminated by the bright and narrow 
nebular \heiia\ line which is very common in Seyfert 2 spectra. 
It is thus very difficult to derive an accurate number of 
WR stars from the broad \heiia\ component. Instead, it is better 
to use the ``red'' WR features around 5800 \AA\ because 
these broad lines are free of any contamination by close nebular 
emission lines. 

\begin{table}
\caption{WR features in nearby Seyfert 2 galaxies}
{\footnotesize
\begin{tabular}{lccl}
\hline
\hline
galaxy & morphology & d (Mpc) & WR features \\
\hline
NGC 6764 & SBbc & 35 & \niii, \heiia, \ciii, and \civ \\
NGC 5427 & SBc & 35 & \niii, \heiia (narrow), and \civ? \\
NGC 4388 & Sb & 33 & \heiia (narrow), and \civ? \\
NGC 5953 & S0-a & 26 & ... \\
NGC 6951 & SABbc & 19 & ... \\
\hline
\hline
\end{tabular}
}
\label{SEY2}
\end{table}

The detection of WR spectral features is more ambiguous in the 
Seyfert 2 galaxies NGC 5427 (see Fig.~\ref{SPEC}) and NGC 4388. 
The blue part of their spectrum around 4700 \AA\ is dominated by 
the bright nebular \heiia\ emission line and the S/N 
around 5800 \AA\ is too low to clearly identify the 
broad \civ\ line. No WR features have been found in the 
two remaining galaxies, NGC 5953 and NGC 6951.

A complete reduction and analysis of the bi-dimensional 
spectra will give us crucial informations on the 
AGN-starburst connection in these galaxies. 
By producing a map of the WR spectral features, 
we should be able (1) to distinguish between a nuclear starburst 
and circumnuclear star-forming regions and (2) to estimate the 
number of massive stars and their contribution to the ionizing and 
bolometric luminosity of the AGN.
 
\section{Conclusions}
WR stars are excellent 
tracers to follow up in relation with the AGN-starburst phenomenon, 
mainly because they provide conspicuous signatures in the visible.
Bi-dimensional spectroscopic observations are currently performed, with 
preliminary results illustrating the feasibility of such a program. 
Spectral features from both WN and WC stars are clearly seen in 
the nucleus of the Seyfert 2 galaxy NGC 6764. An important result is 
the detection of the \ciii\ broad emission line originating from late-type 
WC stars, which are rarely observed in WR galaxies but are expected to be
present in high-metallicity environment, such as an AGN.

\end{document}